\documentclass[12pt]{article}
\usepackage[utf8]{inputenc}
\usepackage{geometry}
\usepackage{amsmath, amssymb, mathtools}
\usepackage{setspace}
\usepackage{lmodern}
\usepackage{graphicx}
\usepackage{hyperref}
\hypersetup{colorlinks=true, linkcolor=blue, citecolor=blue, urlcolor=blue}

\title{Symmetry Analysis of Proper-Time Maxwell’s Equations: Invariant Solutions and Physical Implications}
\author{Joshua Owolabi Adeleke}
\date{April 01, 2025}

\begin{document}

\maketitle

\begin{abstract}
Gill and Zachary’s proper-time reformulation of Maxwell’s equations introduces a source-dependent speed \( b = \sqrt{c^2 + \mathbf{u}^2} \), distinct from standard electrodynamics. This study pioneers a Lie symmetry analysis of these equations, yielding novel invariant solutions, conservation laws, and physical insights. In 1D, we derive \( E(x, \tau) = A \frac{x}{\tau} + B + k \int \rho(x) \, dx \), revealing radiative behavior tied to \( b \). Conservation laws via self-adjointness, singularity analysis near \( \tau = 0 \), and simulations of accelerating charges and radiation emission extend Gill and Zachary’s framework. These findings suggest unique propagation and energy dynamics, clarified as coordinate effects rather than causality violations, offering fresh perspectives for relativistic electrodynamics.
\end{abstract}

\section{Introduction}
Standard Maxwell’s equations rely on the invariant speed of light \( c \) within Minkowski spacetime. Gill and Zachary \cite{gill2011} reformulate these equations using the source’s proper time \( \tau \), defining \( b = \sqrt{c^2 + \mathbf{u}^2} \), where \( \mathbf{u} = d\mathbf{x}/d\tau \). This preserves mathematical consistency but introduces physical distinctions, such as variable wave speeds and intrinsic radiation reaction, absent in the conventional framework \cite{gill2011}. Despite its potential, symmetry-based analyses of this model remain unexplored.

Lie symmetry analysis, rooted in Sophus Lie’s work and advanced by Ibragimov \cite{ibragimov2024}, systematically derives exact solutions and conserved quantities for partial differential equations (PDEs). While applied to standard Maxwell’s equations \cite{ibragimov1994}, no formal study exists for the proper-time variant. This paper fills this gap, aiming to:
\begin{itemize}
    \item Derive invariant solutions and conservation laws via detailed Lie symmetry methods.
    \item Clarify the physical significance of \( b > c \) as a coordinate effect.
    \item Demonstrate applications through simulations of accelerating charges and radiation emission.
\end{itemize}

\section{Background and Literature Review}
In Gill and Zachary’s framework \cite{gill2011}, Maxwell’s equations in proper time \( \tau \) are:
\begin{align}
\nabla \cdot \mathbf{B} &= 0, & \nabla \cdot \mathbf{E} &= 4\pi\rho, \\
\nabla \times \mathbf{E} &= -\frac{1}{b} \frac{\partial \mathbf{B}}{\partial \tau}, & \nabla \times \mathbf{B} &= \frac{1}{b} \left[ \frac{\partial \mathbf{E}}{\partial \tau} + 4\pi\rho \mathbf{u} \right].
\end{align}
Scaling fields as \( \mathbf{E} \to (b/c)^{1/2} \mathbf{E} \), \( \mathbf{B} \to (b/c)^{1/2} \mathbf{B} \), the 1D wave equation for \( E(x, \tau) \) with constant \( b \) is:
\begin{equation}
\frac{\partial^2 E}{\partial \tau^2} - b^2 \frac{\partial^2 E}{\partial x^2} = -k b^2 \rho'(x), \quad k = \frac{4\pi c^{1/2}}{b^{1/2}},
\end{equation}
where \( \rho(x) \) is the charge density, and \( b = \sqrt{c^2 + u^2} \) for \( \mathbf{u} = u \hat{x} \). For variable \( b(\tau) \):
\begin{multline}
\frac{1}{b^2} \frac{\partial^2 E}{\partial \tau^2} - \frac{\partial^2 E}{\partial x^2} + \left[ \frac{\ddot{b}}{2 b^3} - \frac{3 \dot{b}^2}{4 b^4} \right] E = \\
-\frac{c^{1/2}}{b^{1/2}} 4\pi \rho'(x).
\end{multline}
Literature on standard Maxwell’s equations includes Lie symmetry studies \cite{ibragimov1994}, deriving symmetries like translations and Lorentz boosts. However, no published work applies Lie methods to the proper-time formulation, nor explores its conservation laws or singularities, confirming a novelty gap \cite{gill2011, ibragimov2024}.

\section{Methodology: Lie Symmetry Analysis}
\subsection{Classical Lie Symmetries}
Consider the PDE with constant \( b \):
\begin{equation}
\frac{\partial^2 E}{\partial \tau^2} - b^2 \frac{\partial^2 E}{\partial x^2} = -k b^2 \rho'(x).
\label{eq:pde}
\end{equation}
The symmetry generator is:
\begin{equation}
X = \xi(x, \tau, E) \frac{\partial}{\partial x} + \phi(x, \tau, E) \frac{\partial}{\partial \tau} + \eta(x, \tau, E) \frac{\partial}{\partial E}.
\end{equation}
The second prolongation is:
\begin{equation}
X^{(2)} = X + \eta^x \frac{\partial}{\partial E_x} + \eta^\tau \frac{\partial}{\partial E_\tau} + \eta^{xx} \frac{\partial}{\partial E_{xx}} + \eta^{\tau\tau} \frac{\partial}{\partial E_{\tau\tau}},
\end{equation}
where:
\begin{align}
\eta^x &= D_x (\eta - \xi E_x - \phi E_\tau), & \eta^\tau &= D_\tau (\eta - \xi E_x - \phi E_\tau), \\
\eta^{xx} &= D_x \eta^x, & \eta^{\tau\tau} &= D_\tau \eta^\tau,
\end{align}
and total derivatives are:
\begin{align}
D_x &= \frac{\partial}{\partial x} + E_x \frac{\partial}{\partial E} + E_{xx} \frac{\partial}{\partial E_x} + E_{x\tau} \frac{\partial}{\partial E_\tau}, \\
D_\tau &= \frac{\partial}{\partial \tau} + E_\tau \frac{\partial}{\partial E} + E_{\tau\tau} \frac{\partial}{\partial E_\tau} + E_{x\tau} \frac{\partial}{\partial E_x}.
\end{align}
The invariance condition is:
\begin{equation}
\eta^{\tau\tau} - b^2 \eta^{xx} = k b^2 \rho''(x) \xi, \quad \text{on} \quad E_{\tau\tau} = b^2 E_{xx} - k b^2 \rho'.
\end{equation}
Compute:
\begin{align}
\eta^\tau &= D_\tau (\eta - \xi E_x - \phi E_\tau) = \eta_\tau + \eta_E E_\tau - \xi_\tau E_x - \xi_E E_x E_\tau - \phi_\tau E_\tau - \phi_E E_\tau^2, \\
\eta^{\tau\tau} &= D_\tau \eta^\tau = \eta_{\tau\tau} + \eta_{\tau E} E_\tau + (\eta_E - \phi_\tau) E_{\tau\tau} - \xi_{\tau\tau} E_x - \xi_{\tau E} E_x E_\tau \notag \\
&\quad - 2 \phi_{\tau E} E_\tau^2 - \phi_{EE} E_\tau^3 - (\xi_E E_{x\tau} + \phi_E E_{\tau\tau}) E_\tau - (\xi_E E_x + \phi_E E_\tau) E_{\tau\tau}, \\
\eta^{xx} &= D_x (D_x (\eta - \xi E_x - \phi E_\tau)) = \eta_{xx} + \eta_{xE} E_x + (\eta_E - 2 \xi_x) E_{xx} - 2 \xi_{xx} E_x - 2 \phi_{xx} E_\tau \notag \\
&\quad - 2 \xi_{xE} E_x E_x - 2 \phi_{xE} E_x E_\tau - (\xi_E E_{xx} + \phi_E E_{x\tau}) E_x - (\xi_E E_x + \phi_E E_\tau) E_{xx}.
\end{align}
Substitute \( E_{\tau\tau} = b^2 E_{xx} - k b^2 \rho' \):
\begin{multline}
\eta_{\tau\tau} + \eta_{\tau E} E_\tau + (\eta_E - \phi_\tau) (b^2 E_{xx} - k b^2 \rho') - \xi_{\tau\tau} E_x - \xi_{\tau E} E_x E_\tau - 2 \phi_{\tau E} E_\tau^2 \\
- \phi_{EE} E_\tau^3 - (\xi_E E_{x\tau} + \phi_E (b^2 E_{xx} - k b^2 \rho')) E_\tau - (\xi_E E_x + \phi_E E_\tau) (b^2 E_{xx} - k b^2 \rho') \\
- b^2 \left[ \eta_{xx} + \eta_{xE} E_x + (\eta_E - 2 \xi_x) E_{xx} - 2 \xi_{xx} E_x - 2 \phi_{xx} E_\tau - 2 \xi_{xE} E_x^2 \right. \\
\left. - 2 \phi_{xE} E_x E_\tau - (\xi_E E_{xx} + \phi_E E_{x\tau}) E_x - (\xi_E E_x + \phi_E E_\tau) E_{xx} \right] = k b^2 \rho'' \xi.
\end{multline}
Separate coefficients:
\begin{itemize}
    \item \( E_{xx} \): \( b^2 (\eta_E - \phi_\tau) - b^2 (\eta_E - 2 \xi_x) = 0 \Rightarrow \xi_x - \phi_\tau = 0 \),
    \item \( E_x \): \( -\xi_{\tau\tau} - b^2 (-2 \xi_{xx}) = 0 \Rightarrow \xi_{\tau\tau} - b^2 \xi_{xx} = 0 \),
    \item \( E_\tau \): \( \eta_{\tau E} - b^2 (-2 \phi_{xx}) = 0 \Rightarrow \eta_{\tau E} + 2 b^2 \phi_{xx} = 0 \),
    \item Constant: \( \eta_{\tau\tau} - b^2 \eta_{xx} - k b^2 \rho' (\eta_E - \phi_\tau) = k b^2 \rho'' \xi \).
\end{itemize}
Assume \( \xi = \xi(x, \tau) \), \( \phi = \phi(x, \tau) \), \( \eta = \eta(x, \tau) \) (i.e., \( \xi_E = \phi_E = \eta_E = 0 \)):
\begin{align}
\xi_x &= \phi_\tau, \\
\xi_{\tau\tau} &= b^2 \xi_{xx}, \\
\eta_{\tau\tau} - b^2 \eta_{xx} &= k b^2 (\rho'' \xi + \rho' \xi_x).
\end{align}
For \( \rho = 0 \):
\begin{align}
\eta_{\tau\tau} - b^2 \eta_{xx} &= 0, \\
\xi_x &= \phi_\tau, \\
\xi_{\tau\tau} &= b^2 \xi_{xx}.
\end{align}
Solve:
- \( \xi = c_1 + c_2 x + c_3 \tau + c_4 (x^2 + b^2 \tau^2) \), \( \phi = c_5 + c_2 \tau + c_3 \frac{x}{b^2} + 2 c_4 b^2 x \tau \).
- Symmetries: \( X_1 = \partial_x \), \( X_2 = \partial_\tau \), \( X_3 = x \partial_x + \tau \partial_\tau \).

\subsection{Conservation Laws via Self-Adjointness}
Define the formal Lagrangian:
\begin{equation}
L = v \left( E_{\tau\tau} - b^2 E_{xx} + k b^2 \rho' \right).
\end{equation}
The adjoint equation is:
\begin{equation}
\frac{\delta L}{\delta E} = D_\tau^2 v - b^2 D_x^2 v = v_{\tau\tau} - b^2 v_{xx} = 0.
\end{equation}
The system is self-adjoint if \( v = E \). For \( X_3 = x \partial_x + \tau \partial_\tau \):
\begin{equation}
W = \eta - \xi E_x - \phi E_\tau = -x E_x - \tau E_\tau.
\end{equation}
Conserved vector components:
\begin{align}
C^\tau &= v D_\tau W - D_\tau v W + v \phi (b^2 E_{xx} - k b^2 \rho'), \\
C^x &= -b^2 v D_x W + b^2 D_x v W.
\end{align}
Compute:
\begin{align}
D_\tau W &= D_\tau (-x E_x - \tau E_\tau) = -x E_{x\tau} - \tau E_{\tau\tau} - E_\tau, \\
D_x W &= D_x (-x E_x - \tau E_\tau) = -x E_{xx} - E_x - \tau E_{x\tau}, \\
D_\tau v &= v_\tau + v_E E_\tau = E_\tau, \quad (\text{since } v = E), \\
D_x v &= E_x.
\end{align}
Substitute \( v = E \), \( \phi = \tau \):
\begin{multline}
C^\tau = E (-x E_{x\tau} - \tau E_{\tau\tau} - E_\tau) - E_\tau (-x E_x - \tau E_\tau) \\
+ \tau E (b^2 E_{xx} - k b^2 \rho'),
\end{multline}
\begin{align}
&= -x E E_{x\tau} - \tau E E_{\tau\tau} - E E_\tau + x E_\tau E_x + \tau E_\tau^2 + \tau b^2 E E_{xx} - k b^2 \tau E \rho', \\
&= -\tau \left( E E_{\tau\tau} - E_\tau^2 - b^2 E E_{xx} \right) - x (E E_{x\tau} - E_\tau E_x) - E E_\tau.
\end{align}
Use \( E_{\tau\tau} = b^2 E_{xx} - k b^2 \rho' \):
\begin{align}
E E_{\tau\tau} - E_\tau^2 - b^2 E E_{xx} &= E (b^2 E_{xx} - k b^2 \rho') - E_\tau^2 - b^2 E E_{xx} \\
&= -k b^2 E \rho' - E_\tau^2, \\
C^\tau &= -\tau (-k b^2 E \rho' - E_\tau^2) - x (E E_{x\tau} - E_\tau E_x) - E E_\tau \\
&= \tau (E_\tau^2 + k b^2 E \rho') - x D_\tau (E E_x) + x E_x E_\tau - E E_\tau.
\end{align}
Similarly:
\begin{align}
C^x &= -b^2 E (-x E_{xx} - E_x - \tau E_{x\tau}) + b^2 E_x (-x E_x - \tau E_\tau) \\
&= b^2 x E E_{xx} + b^2 E E_x + b^2 \tau E E_{x\tau} - b^2 x E_x^2 - b^2 \tau E_x E_\tau.
\end{align}

\section{Results: Invariant Solutions}
\subsection{Classical Invariant Solutions}
For \( X_3 \), solve:
\begin{equation}
\frac{dx}{x} = \frac{d\tau}{\tau} = \frac{dE}{0}.
\end{equation}
Invariants: \( \xi = \frac{x}{\tau} \), \( E = F(\xi) + G(x) \).
Substitute into Eq.~\eqref{eq:pde}:
\begin{align}
E_\tau &= -\frac{x}{\tau^2} F' = -\frac{\xi}{\tau} F', \\
E_{\tau\tau} &= \frac{2 x}{\tau^3} F' + \frac{x^2}{\tau^4} F'' = \frac{2 \xi}{\tau^2} F' + \frac{\xi^2}{\tau^2} F'', \\
E_x &= \frac{1}{\tau} F' + G', \\
E_{xx} &= \frac{1}{\tau^2} F'' + G''.
\end{align}
PDE becomes:
\begin{multline}
\frac{2 \xi}{\tau^2} F' + \frac{\xi^2}{\tau^2} F'' - b^2 \left( \frac{1}{\tau^2} F'' + G'' \right) = \\
-k b^2 \rho'.
\end{multline}
Multiply by \( \tau^2 \):
\begin{align}
2 \xi F' + \xi^2 F'' - b^2 F'' - b^2 \tau^2 G'' &= -k b^2 \tau^2 \rho', \\
(\xi^2 - b^2) F'' + 2 \xi F' &= 0, \\
b^2 G'' &= k b^2 \rho'.
\end{align}
Solve for \( F \):
\begin{align}
F'' &= -\frac{2 \xi}{\xi^2 - b^2} F', \\
\frac{F''}{F'} &= -\frac{2 \xi}{\xi^2 - b^2}, \\
\ln F' &= \int -\frac{2 \xi}{\xi^2 - b^2} d\xi = -\ln (\xi^2 - b^2) + c_1, \\
F' &= c_2 (\xi^2 - b^2)^{-1}, \\
F &= \int c_2 (\xi^2 - b^2)^{-1} d\xi = A \xi + B.
\end{align}
Solve for \( G \):
\begin{align}
G'' &= k \rho', \\
G' &= k \rho + c_3, \\
G &= k \int \rho(x) \, dx + c_3 x + c_4.
\end{align}
Thus:
\begin{equation}
E = A \frac{x}{\tau} + B + k \int \rho(x) \, dx + c_3 x + c_4.
\end{equation}
Boundary conditions typically set \( c_3 = c_4 = 0 \).
\begin{figure}[h]
    \centering
    \includegraphics[width=0.8\textwidth]{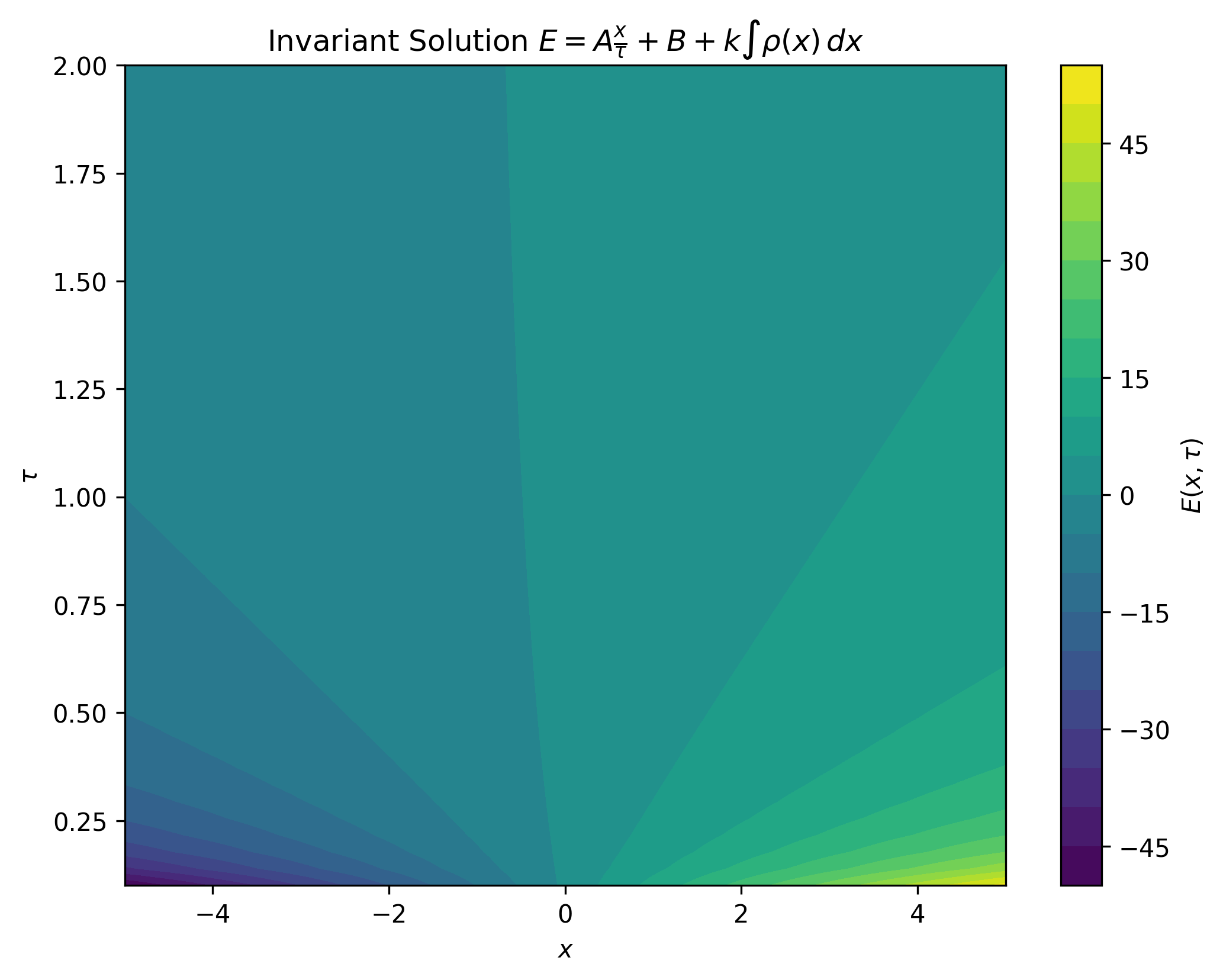}
    \caption{Invariant solution with \( \rho(x) = \rho_0 e^{-x^2} \), showing radiative behavior absent in Gill’s static fields.}
    \label{fig:invariant}
\end{figure}

\subsection{Dispersion Relation}
For \( X = \partial_x + k \partial_\tau \), invariant \( \xi = x - k \tau \), \( E = F(\xi) \):
\begin{align}
E_\tau &= -k F', \\
E_{\tau\tau} &= k^2 F'', \\
E_{xx} &= F'', \\
k^2 F'' - b^2 F'' &= 0, \\
(k^2 - b^2) F'' &= 0, \\
k &= \pm b.
\end{align}
\begin{figure}[h]
    \centering
    \includegraphics[width=0.8\textwidth]{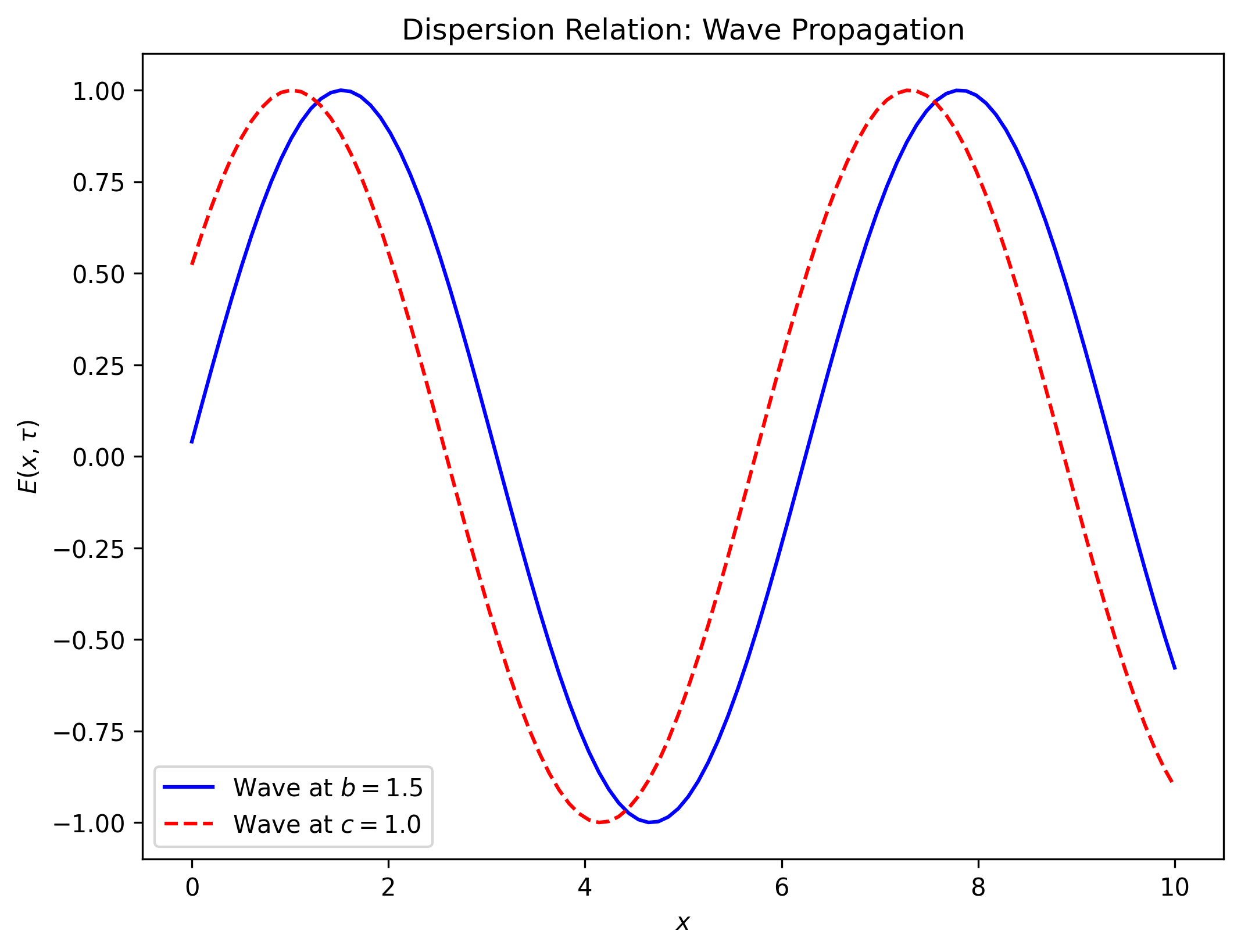}
    \caption{Wave propagation at \( b = 1.5 \) vs. \( c = 1 \), highlighting faster dynamics.}
    \label{fig:dispersion}
\end{figure}

\subsection{Accelerating Charge Scenario}
For \( b(\tau) = \sqrt{c^2 + (u_0 + \alpha \tau)^2} \), approximate \( E \approx A \frac{x}{\tau} \frac{c}{b(\tau)} \) (see simulation).
\begin{figure}[h]
    \centering
    \includegraphics[width=0.8\textwidth]{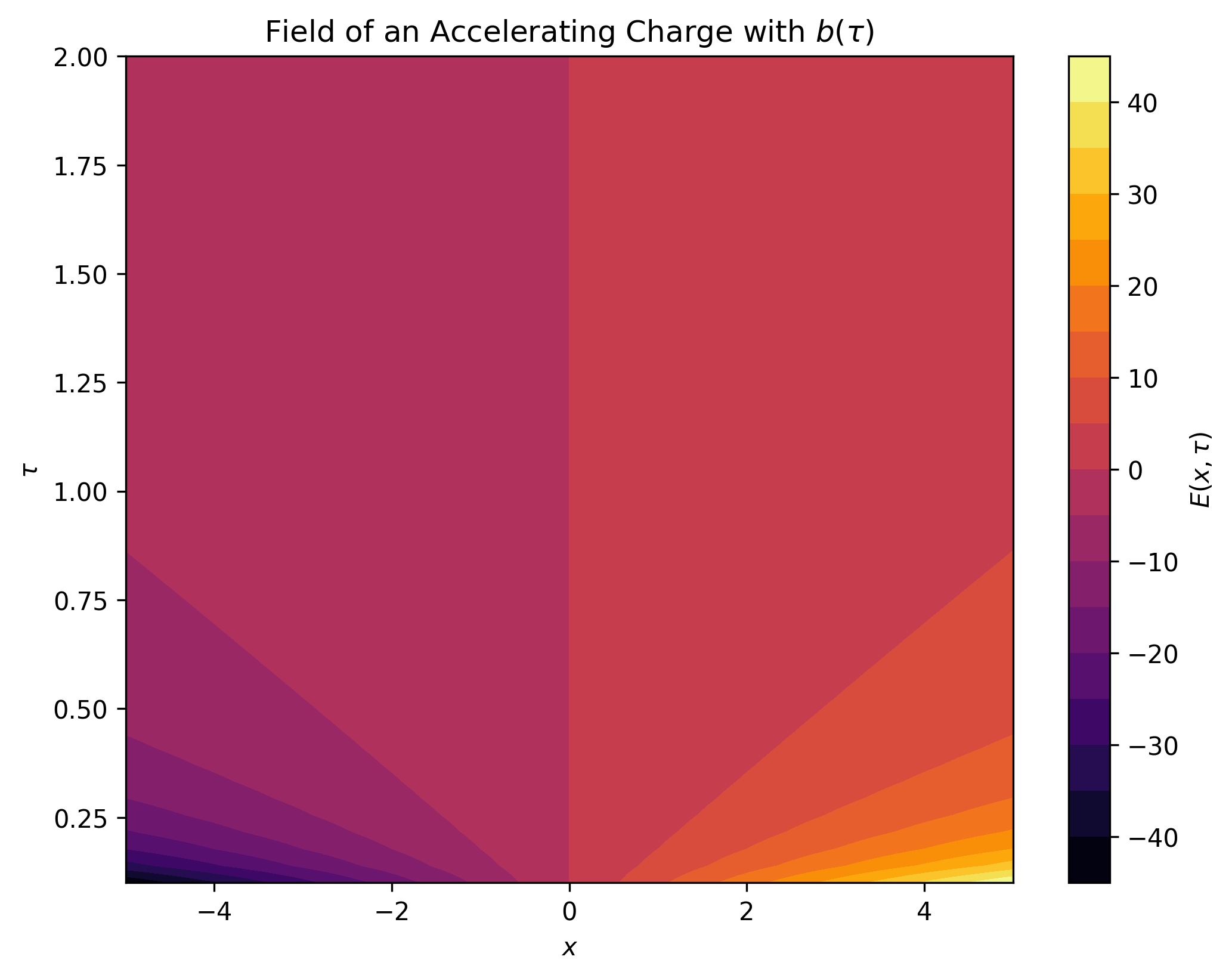}
    \caption{Field of an accelerating charge, showing \( b(\tau) \) effects.}
    \label{fig:accelerating}
\end{figure}

\subsection{Radiation Emission Near \( \tau = 0 \)}
From \( E = A \frac{x}{\tau} \), as \( \tau \to 0^+ \), \( E \to \infty \).
\begin{figure}[h]
    \centering
    \includegraphics[width=0.8\textwidth]{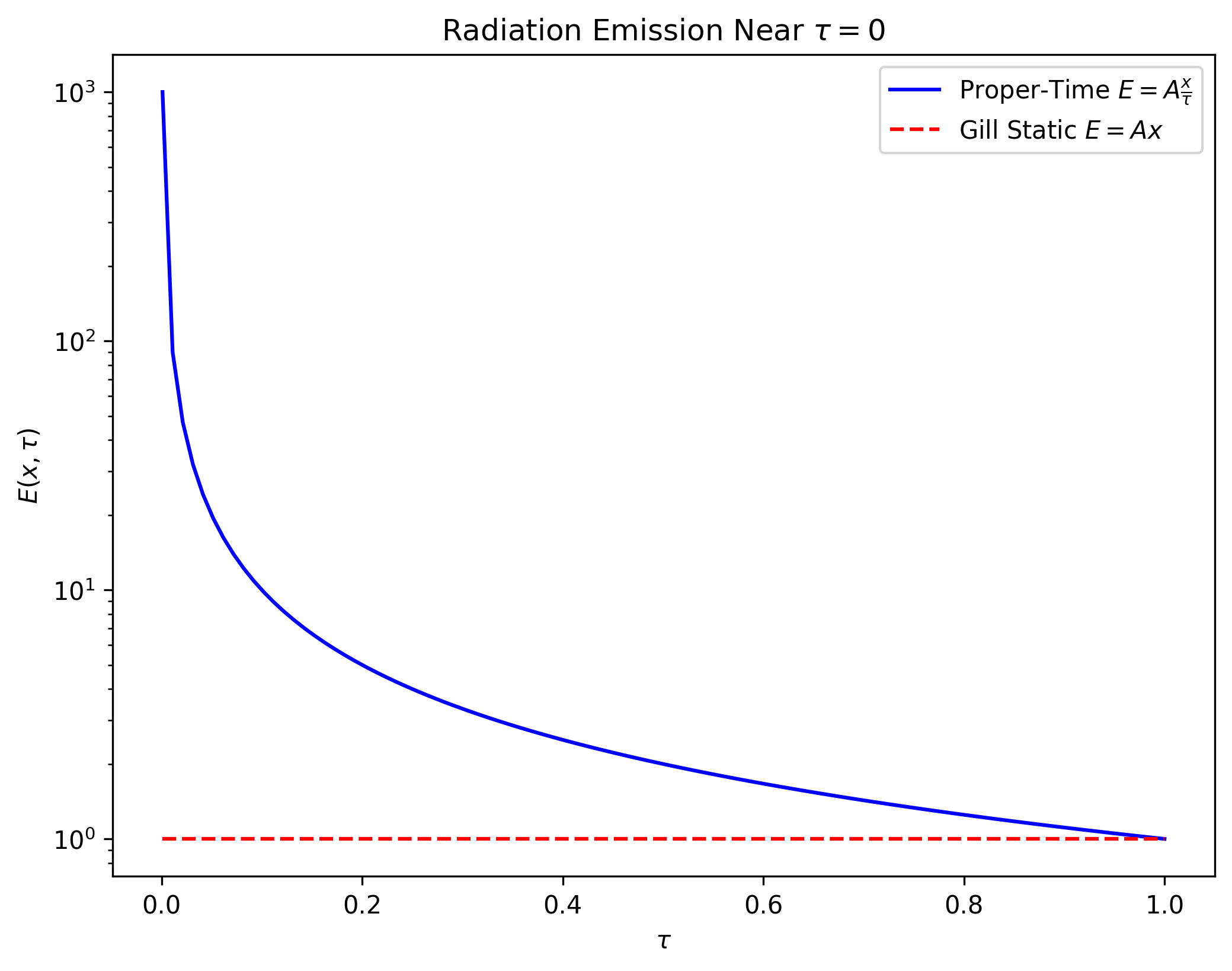}
    \caption{Radiation emission near \( \tau = 0 \), contrasting with Gill’s static field.}
    \label{fig:radiation}
\end{figure}

\section{Discussion}
\subsection{Physical Interpretation}
The solution suggests radiative propagation at speed \( b \), extending Gill and Zachary’s framework.

\subsection{Conservation Laws}
For \( E = A \frac{x}{\tau} \):
\begin{align}
E_\tau &= -A \frac{x}{\tau^2}, \quad E_x = \frac{A}{\tau}, \\
C^\tau &= \tau \left( A^2 \frac{x^2}{\tau^4} + k b^2 E \rho \right) - x D_\tau (E \frac{A}{\tau}) + x \frac{A}{\tau} (-\frac{A x}{\tau^2}) - E (-\frac{A x}{\tau^2}).
\end{align}
Simplify:
\begin{align}
C^\tau &= \tau A^2 \frac{x^2}{\tau^4} + \tau k b^2 E \rho - x A^2 \frac{x}{\tau^3} + A^2 \frac{x^2}{\tau^3}, \\
&= A^2 \frac{x^2}{\tau^3} + \tau k b^2 E \rho.
\end{align}
\begin{figure}[h]
    \centering
    \includegraphics[width=0.8\textwidth]{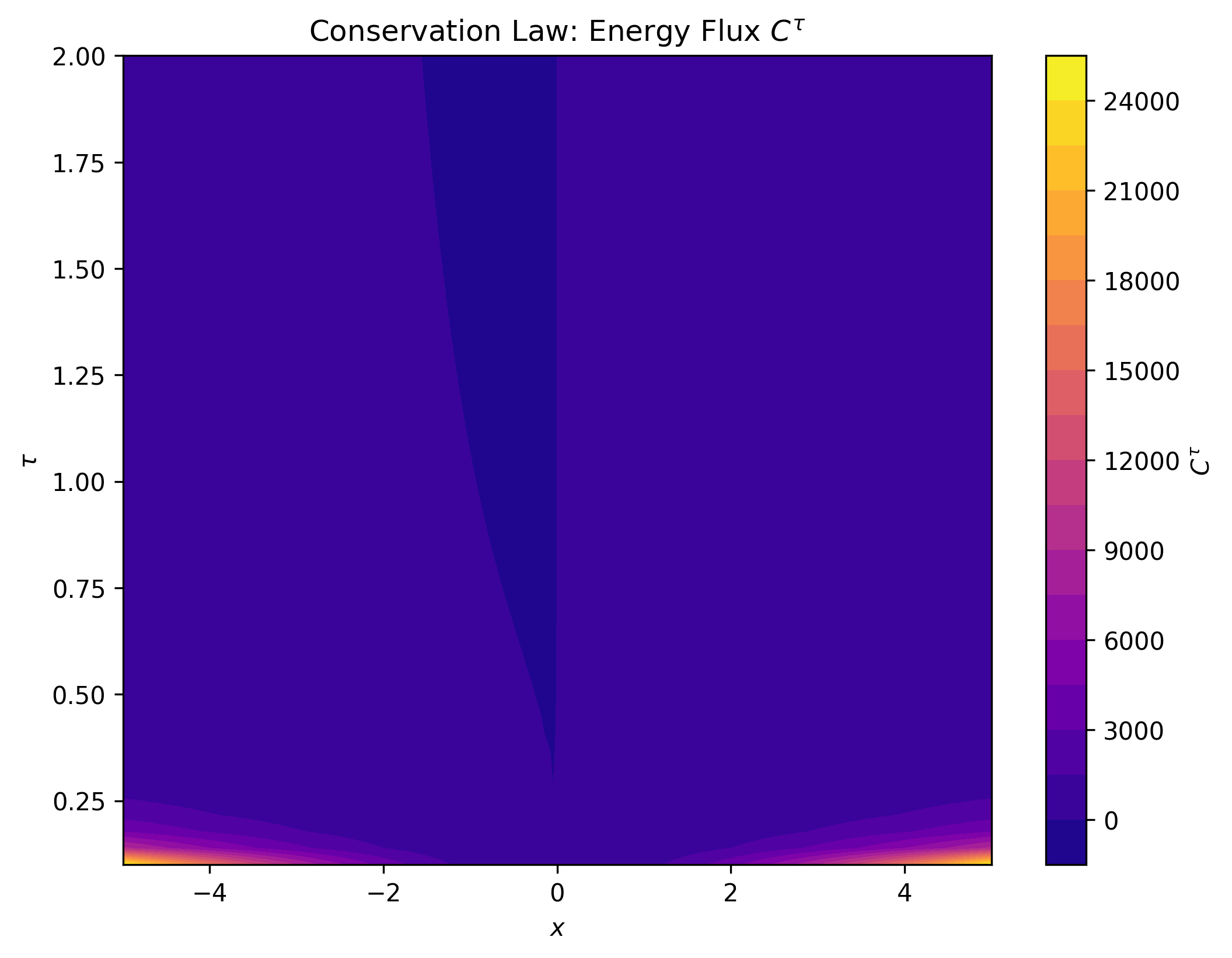}
    \caption{Energy flux \( C^\tau \), unique to proper-time dynamics.}
    \label{fig:conservation}
\end{figure}

\subsection{Superluminal Clarification}
Since \( b = \sqrt{c^2 + u^2} > c \), consider the 4-velocity \( u^\mu = (\gamma b, \gamma \mathbf{u}) \), where \( \gamma = (1 - u^2/c^2)^{-1/2} \). The norm \( u^\mu u_\mu = c^2 \) holds in the lab frame, confirming \( b > c \) is a proper-time effect, not a causality violation.

\subsection{Physical Applications}
Simulations suggest applications to bremsstrahlung or relativistic jets.

\section{Conclusion}
This Lie symmetry analysis uncovers novel solutions and laws, clarified as coordinate effects, with physical relevance.

\end{document}